\documentclass[final,3p,12pt]{elsarticle}

\usepackage{amssymb}
\usepackage{amsmath}
\usepackage{hyperref}
\journal{Current Opinion in Structural Biology}

\begin{document}

\begin{frontmatter}

\title{Simulation-based inference of single-molecule experiments}

\author[inst1,inst2]{Lars Dingeldein}
\author[inst3,inst4]{Pilar Cossio}
\author[inst5,inst2]{Roberto Covino}

\affiliation[inst1]{organization={Institute of Physics, Goethe University Frankfurt},%Department and Organization 
            city={Frankfurt am Main},
            country={Germany}}

\affiliation[inst2]{organization={Frankfurt Institute for Advanced Studies},%Department and Organization
            city={Frankfurt am Main},
            country={Germany}}

\affiliation[inst3]{organization={Center for Computational Mathematics , Flatiron Institute},%Department and Organization
            city={New York},
            country={United States}}

\affiliation[inst4]{organization={Center for Computational Biology, Flatiron Institute},%Department and Organization
            city={New York},
            country={United States}}

\affiliation[inst5]{organization={Institute of Computer Science, Goethe University Frankfurt},%Department and Organization
            city={Frankfurt am Main},
            country={Germany}}

\begin{abstract}
Single-molecule experiments are a unique tool to characterize the structural dynamics of biomolecules. However, reconstructing molecular details from noisy single-molecule data is challenging. 
Simulation-based inference (SBI) integrates statistical inference, physics-based simulators, and machine learning and is emerging as a powerful framework for analysing complex experimental data. 
Recent advances in deep learning have accelerated the development of new SBI methods, enabling the application of Bayesian inference to an ever-increasing number of scientific problems.
Here, we review the nascent application of SBI to the analysis of single-molecule experiments. We introduce parametric Bayesian inference and discuss its limitations. We then overview emerging deep-learning-based SBI methods to perform Bayesian inference for complex models encoded in computer simulators. We illustrate the first applications of SBI to single-molecule force-spectroscopy and cryo-electron microscopy experiments. 
SBI allows us to leverage powerful computer algorithms modeling complex biomolecular phenomena to connect scientific models and experiments in a principled way.    
\end{abstract}

\begin{keyword}
%% keywords here, in the form: keyword \sep keyword
simulation-based inference \sep likelihood-free inference \sep single-molecule data analysis \sep Bayesian inference \sep data analysis
\end{keyword}

\end{frontmatter}

\section{Introduction}
\label{sec:introduction}
The structural dynamics of biomolecules and their assemblies determine their functions. 
For instance, knowing how a protein reorganizes between alternative conformations, the mechanism, thermodynamics, and kinetics of the process, is key to understanding its function. Most traditional biophysical experiments report ensemble measurements,  where an observable is averaged over all conformations weighted by their frequency. While invaluable, these experiments are inadequate to characterize the inherent heterogeneity and stochasticity of biomolecules.

\begin{figure*}[t]
    \centering
    \includegraphics[width=0.99\textwidth]{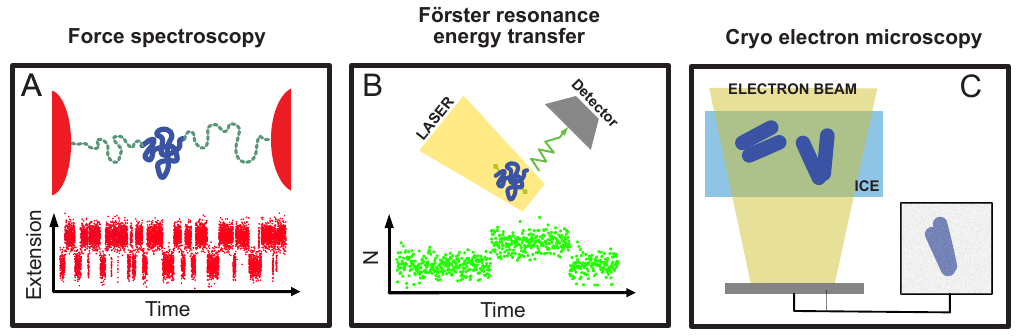}
 e    \caption{Schematics of different single-molecule experiments: (A) Force spectroscopy measures the distance between devices (red) attached via flexible linkers (green) to an individual biomolecule (blue) over time. (B) Förster resonance energy transfer measures the distance between two fluorescence dyes. (C) Cryo electron Microscopy images ensembles of molecules in a frozen solution.}
    \label{fig:1}
\end{figure*}

Single-molecule experiments probe individual biomolecules, enabling the statistical characterization of their structural dynamics.  
Techniques like single-molecule force spectroscopy (smFS)\cite{woodside2014reconstructing} (Fig. \ref{fig:1}A) and single-molecule Förster resonance energy transfer (smFRET)\cite{nettels2024single} (Fig. \ref{fig:1}B)  measure time series of distances between pairs of specific sites. Cryo-electron microscopy (cryo-EM) is also a potential single-molecule technique. While state-of-the-art methods enable only the reconstruction of a handful of alternative conformational states, in principle, cryo-EM produces data that are 2-dimensional snapshots of many identical copies of the same molecule in all the possible conformations \cite{tang2023conformational} (Fig. \ref{fig:1}C). 

Reconstructing biomolecular structural dynamics from sparse and noisy single-molecule measurements is an ill-posed problem. For example, while describing the dynamics of a protein requires specifying the trajectories of all its atoms, smFS and smFRET report a handful of distances. Reconstructing the structural dynamics is impossible unless we make strong prior assumptions.

Statistical inference provides a principled framework to learn biomolecular mechanisms, thermodynamics, and kinetics from noisy single-molecule data \cite{presse_sgouralis_2023}. While non-parametric inference is a powerful emergent tool \cite{kilic2023gene}, we focus on parametric inference, where we postulate a biophysical model and learn the value of its parameters from the data. 

Here, we will review simulation-based inference (SBI) \cite{cranmer2020frontier}, an innovative merger of machine learning, statistical inference, and physics-based simulators. SBI is already a key methodology in fields ranging from astrophysics to particle physics and neuroscience and  is emerging as a powerful tool for the analysis of single-molecule experiments. 

\section{Statistical inference}
\label{sec: statistical inference}
The goal of parametric inference is to learn the parameter distributions from experimental data given a model. We describe the data-generation process of experimental data $x^{\mathrm{exp}}$ with the mathematical model $x=\mathcal{M}(\theta)$ with parameters $\theta$. If the model and parameters are accurate, $x$ and $x^{\mathrm{exp}}$ should be similar enough. For example, $x^{\mathrm{exp}}_t$ could be a telegraph-like time series measured in smFS characterizing the repeated folding and unfolding of a protein (Fig. \ref{fig:1}A). A simple model of this measurement could be a one-dimensional Brownian particle on a potential, with $\theta$ quantifying the shape of the potential and the diffusion constant. The synthetic data $x_t$ would be a Brownian trajectory, ideally very similar to $x^{\mathrm{exp}}_t$.  

Bayesian statistics is a general framework for parameter inference. The main output is the posterior $p(\theta|x)$, the probability distribution of $\theta$ conditioned on some data $x$. The posterior is computed using Bayes' Theorem
\begin{equation}
    p(\theta|x) = \frac{p(x|\theta)\, p(\theta)}{p(x)}~.
    \label{eq:posterior}
\end{equation}
Here, $p(x|\theta)$ is the likelihood, the probability of generating the data $x$ given the model and $\theta$. The prior $p(\theta)$ quantifies all pre-existing knowledge about the parameters, and $p(x)=\int p(x| \theta) \, p(\theta) \mathrm{d}\theta$ is the  normalization constant.

\subsection{Maximum Likelihood and Bayesian Inference of single-molecule experiments}

Maximum likelihood estimation (MLE) is a powerful framework often more tractable than Bayesian inference. Maximizing the likelihood w.r.t. the parameters provides the point estimate $\hat{\theta} = \underset{\theta}{\operatorname{argmax}} \, p(x|\theta)$.
MLE was applied, for example, to infer photon trajectories in smFRET \cite{gopich2023analysis},  reconstruct conformations in cryo-EM \cite{cossio2018likelihood}, and characterize molecular complexes in the crowded cellular environment \cite{lucas2021locating,cruz2024high}. The main limitation of MLE is that it provides a single parameter point estimate, which complicates estimating uncertainties and dealing with multi-modal posterior distributions. 

The application of Bayesian inference to analyze single-molecule biophysical data has been limited. Examples are the reconstruction of structural ensembles from very noisy single-molecule X-ray scattering data \cite{schultze2023novo}, the identification of transitions from single-molecule trajectories  \cite{rojewski2024accurate}, and the analysis of smFRET data \cite{bronson2009learning}. In Cryo-EM, Bayesian methods were successfully applied to refine molecular ensembles \cite{tang2023ensemble,hoff2024accurate}.
Despite being conceptually simple, Bayesian inference is often computationally intractable.  

\section{Simulation-based inference}

The trade-off between a model's simplicity and its accuracy is a central challenge in understanding complex biological phenomena. Models can be like caricatures,  simplified or exaggerated representations allowing deep understanding (the proverbial spherical cow). In biology, small variations can have profound effects, and accurate predictive models are crucial for capturing these nuances.
Prioritizing accuracy requires complex mathematical models.  
Scientific models are shifting from being expressed with equations, to being encoded in computer algorithms performing detailed simulations.   

Molecular simulations can reproduce experiments quantitatively \cite{tesei2024conformational} and provide a detailed interpretation of single-molecule data \cite{topel_learned_2023}.  
MLE and Bayesian inference provide a principled framework to integrate simulation and experimental data \cite{kummerer2021fitting, kofinger2024encoding}. 
While algorithms liberate us from the limitations of analytical tractability, using them for statistical inference is a challenge.

\begin{figure*}[t]
    \centering
    \includegraphics[width=\textwidth]{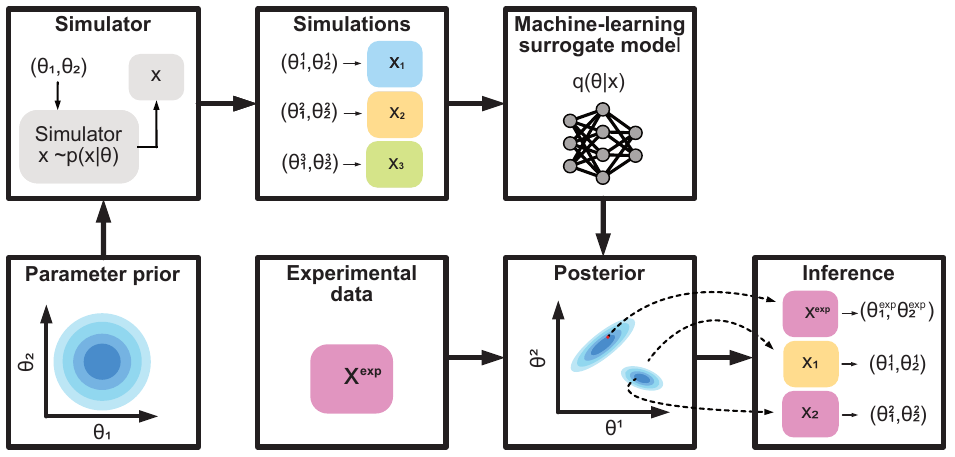}
    \caption{Schematic representation of Neural Posterior Estimation. We aim to explain an experimental observation $x^{\mathrm{exp}}$ using a parametric model encoded in a simulator. Prior parameters $\theta$ are drawn from a prior distribution $p(\theta)$ and used as input for the simulator. The simulator generates synthetic data $x_i$ with parameter $\theta_i$. The data set of parameters and simulations are used to train a surrogate model $q(\theta|x)$, that approximates the true posterior. After the training, the posterior can be evaluated using the experimental observation $x^{\mathrm{exp}}$, revealing the region of the parameter space consistent with the experimental data.}
    \label{fig:2}
\end{figure*}

\subsection{The problem of intractable likelihoods}

Often the likelihood of a complex simulator is too expensive to evaluate, or not even known explicitly. 
Computational intractability has two main origins: latent variables and nuisance parameters. 

Experiments only capture a minority of the degrees of freedom of a complex simulator. For instance, smFRET  reports a handful of distances, while a molecular simulation will explicitly reproduce the trajectory of every atom. The corresponding likelihood must sum up the trajectories of all ``latent'' degrees of freedom $z$, which are explicit in the simulator but hidden in the experiment. This corresponds to the marginalization
\begin{equation}
    p(x|\theta) = \int  p(x, z|\theta) \mathrm{d} z.
\end{equation}\label{eq: latent_z}

Models can have many parameters, but usually only a subset $\theta$ is of scientific interest, while the remaining ones, $\phi$, are nuisance parameters necessary for technical reasons. For example, a simulator modeling a microscope requires parameters to describe details of the image formation process, while we are usually only interested in parameters describing the molecular event being probed. Making inference on $\theta$ taking into account all possible values of $\phi$ requires again marginalizing the likelihood
\begin{equation}
	p(x|\theta) = \int p(x|\theta,\phi) p(\phi) \mathrm{d} \phi,
\end{equation}
where $p(\phi)$ is the prior on the nuisance parameters.
Likelihood marginalization corresponds to high dimensional integrals that are, in general, extremely expensive to evaluate explicitly.  

\begin{figure*}[t]
    \centering
    \includegraphics[width=\textwidth]{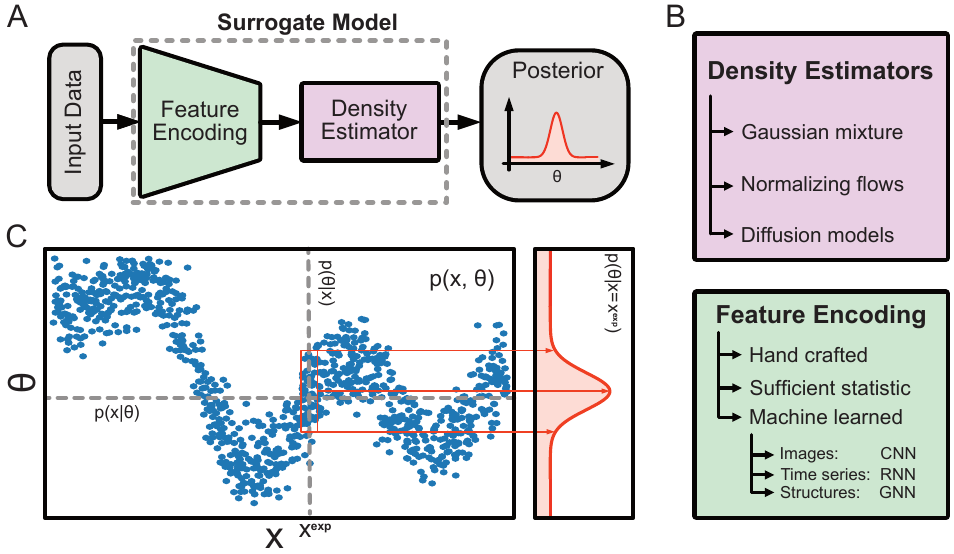}
    \caption{Key ingredients of a posterior surrogate model. (A) The surrogate model of the posterior consists of two parts. The feature encoding (green) compresses the input data into a lower dimensional representation. The density estimator (red) learns a probabilistic mapping between data and parameters. (B) Samples (blue dots) of the joint probability distribution $p(x, \theta)$, the posterior $p(\theta| x)$ (vertical grey line) and the likelihood $p(x| \theta)$ (horizontal grey line) are orthogonal slices through the joint distribution. (C) Examples of different density estimators and feature encoding.}
    \label{fig:3}
\end{figure*}

\subsection{Simulation-based inference in the age of machine learning}

SBI overcomes the challenge of intractable likelihoods by avoiding explicit likelihood or posterior \cite{cranmer2020frontier}. SBI enables the use of complex simulators for Bayesian inference of experimental data. The simulator models the data-generation process, providing a probabilistic mapping between parameters $\theta$ and data $x$ and encoding the likelihood implicitly in the form of an algorithm.  

SBI leverages modern machine learning to learn surrogate models of likelihood and posterior. The surrogates are probabilistic models containing deep neural networks \cite{lueckmann2019likelihood, papamakarios2016fast} approximating the likelihood  \cite{papamakarios2019sequential}, likelihood ratios \cite{hermans2020likelihood}, the posterior \cite{greenberg2019automatic}, or posterior and likelihood simultaneously \cite{gloeckler2024all}. In the following section, we will focus on learning the posterior.

Neural Posterior Estimation is a powerful method to learn a surrogate of the posterior  (Fig. \ref{fig:2}) \cite{gonccalves2020training}. The surrogate $q_{\nu}(\theta|x)$ is a parametric model for conditional density estimation (Fig. \ref{fig:3}A), such as a Gaussian Mixture Model or a Normalizing Flow (Fig. \ref{fig:3}B). The simulator generates a dataset $\mathcal{D} = \{(x_i, \theta_i)\}_{i=1}^N$ (Fig. \ref{fig:2}) containing $N$ simulations $x_i \sim p(x|\theta_i)$ for parameter values $\theta_i \sim p(\theta)$. Learning the posterior requires fitting the surrogate model $q_{\nu}(\theta|x)$ to the data set $\mathcal{D}$ by maximizing $\frac{1}{N}\sum_{i=1}^{N}q_{\nu}(\theta_i|x_i)$. The trained surrogate provides a probabilistic mapping between parameters and data, enabling inference for 
an experimental observation $x^{\mathrm{exp}}$: $p(\theta|x=x^{\mathrm{exp}}) \approx q_{\nu}(\theta|x=x^{\mathrm{exp}})\label{eq1}$ (Fig. \ref{fig:3}C).

Experimental observations $x^{\mathrm{exp}}$ are usually high-dimensional data such as images or time series. Feeding the data to the surrogates requires compressing them into a lower-dimensional representation, ideally without losing crucial information (Fig. \ref{fig:3}A, B). The compression could be done by hand-crafted features, such as  the number of peaks in a time series. Alternatively, by using an adequate number of simple statistical descriptors, such as statistical moments. 
Deep learning provides the tools to learn the features directly from the data. For this, we need an additional embedding network $S_{\psi}$ that compresses the observation $x$ into $v = S_{\psi}(x)$ (Fig. \ref{fig:3}C). The embedding network and the surrogate are trained jointly by maximizing $\frac{1}{N}\sum_{i=0}^{N}q_{\nu}(\theta_i|S_{\psi}(x_i))$.

The posterior surrogate enables amortized inference. The computational cost of simulating and training is paid only once upfront, after which inference requires only an evaluation of the neural network in the surrogate. The negligible cost of doing inference allows for on-the-fly inference or analysis of arbitrarily large data sets. Amortization is a significant advantage compared to the explicit optimization method, where the likelihood must be optimized for every observation. 

\begin{figure*}[t]
    \centering
    \includegraphics[width=\textwidth]{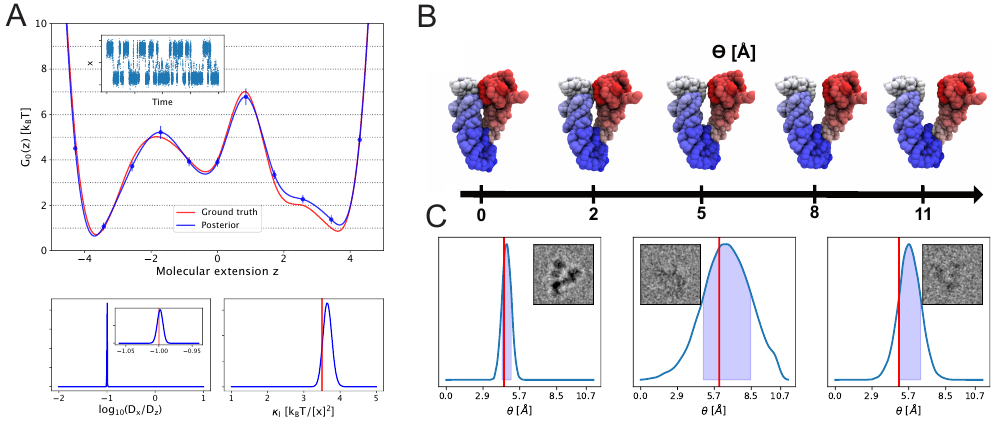}
    \caption{Proof-of-concept SBI applications for (A) Single-molecule force spectroscopy and (B and C) cryo-electron microscopy. (A, Upper panel). Reconstruction of a molecular free energy surface. The red and blue curves show the ground truth and reconstruction, respectively. The inset shows the trajectory used to reconstruct the surface. (Lower panels) Posteriors estimating diffusivities and linker stiffness from the same data. (B) Conformational change along of the model system hsp90, parameterized by $\theta$.
   (C) Example posteriors for synthetic cryo-em images of hsp90 and corresponding posterior distributions. The red line indicates the ground truth conformation, visible in the cryo-em image.}
    \label{fig:4}
\end{figure*}

\subsection{SBI for single-molecule force spectroscopy}

The analysis of smFS was one of the first applications of SBI on single-molecule data \cite{dingeldein2023simulation}. smFS probes conformational changes by mechanically manipulating an individual biomolecule.
A typical constant-force experiment attaches a pulling device to a biomolecule via flexible polymers (Fig. \ref{fig:1}A). The experiment reports the time series of the measured extension $x_t$, which only indirectly reflects the molecular extension $z_t$ while the molecule unfolds and refolds (Fig. \ref{fig:4}A). The measured extension is the outcome of the complex interplay between the large and slow pulling device, the linkers, and the molecule. Naive analyses of $x_t$ lead to severe artifacts \cite{cossio2015artifacts}. 

Simple physical models of smFS result in intractable likelihoods.
The well-established harmonic-linker model \cite{covino2019molecular} describes the molecular extension time series $z_t$ as diffusive on a free-energy surface. The measured extension time series $x_t$ results from a harmonic coupling with $z_t$. Crucially, we can only measure $x_t$, while $z_t$ is latent.  
The likelihood of this simple model has a dozen parameters: the spline node to approximate the molecular energy profile $G(z)$, the ratio of diffusion coefficients $D_x/D_z$, and the linker stiffness $k_l$. 
For this model, Bayes' Theorem becomes
\begin{equation}
    p(\theta|x_t) = \frac{\int p(x_t, z_t|\theta) p(\theta)\mathrm{d}z_t}{p(x_t)},
\end{equation}
where $\theta$ groups all model parameters. 
Remarkably, the likelihood of this very simple model is a path-integral over all possible latent trajectories $z_t$, which is extremely expensive to optimize. 

Overcoming this challenge with NPE is straightforward. Three components are needed: a simulator, a prior, and a density estimator. The simulator, a simple code that numerically integrate Brownian trajectories, implicitly encodes the marginal likelihood. The training data set $\mathcal{D} = \{(\theta_i, x_t^{i})\}_{i=1}^N$ for training the surrogate model is generated by sampling $N$ parameters from the prior $\theta_i \sim p(\theta)$ and simulate the measured extension $x_t^{i} \sim p(x_t|\theta=\theta_i)$. The time series $q_t$ is featurized in a vector $s$ with a set of statistics. A surrogate model $q_{\nu}(\theta|s)$ is then trained on $\mathcal{D} = \{(\theta_i, s_i)\}_{i=1}^N$. The trained surrogate approximates the true posterior $q_{\nu}(\theta|s) \approx p(\theta|x_t)$, and enables accurate inference of the model parameters without any explicit evaluation of the likelihood (Fig. \ref{fig:4}A). 

\subsection{Identifying individual conformations in cryo-EM images with cryoSBI}

Single-particle cryo-EM captures two-dimensional snapshots of individual molecules in different conformations. These images are very noisy projections along unknown angles. Even though cryo-EM data could give direct access to the entire conformational ensemble, identifying the conformation in an image is an outstanding inference problem. 

Tackling this problem with explicit likelihood approaches is computationally very demanding.   
Assuming a set of biomolecular conformations parameterized by $\theta$ (Fig. \ref{fig:4}B) and  a cryo-EM image $I_{\mathrm{obs}}$ containing a single biomolecule in conformation $\theta_{\mathrm{obs}}$ (Fig. \ref{fig:4}C), the goal is to infer the conformation $\theta_{\mathrm{obs}}$ from the image $I_{\mathrm{obs}}$ with Bayes' theorem: 
\begin{equation}
    p(\theta|I_{\mathrm{obs}}) = \frac{\int p(I_{\mathrm{obs}}|\theta, \phi) p(\theta) p(\phi) \mathrm{d}\phi}{p(I_{\mathrm{obs}})}.
\end{equation}
Making inference requires a marginalization over all nuisance parameters $\phi$, describing the unknown projection angle and the physics of the image formation process, such as the microscope's point spread function.  
The explicit marginalization is computationally very challenging \cite{cossio2017bioem}. On the other hand, simulating an image for a given set of parameters is relatively straightforward and computationally inexpensive. 

cryoSBI is a recent proof-of-concept demonstrating the promise of using NPE to identify single conformations from single Cryo-EM images \cite{dingeldein2024amortized}. cryoSBI starts with sampling conformations $\theta_i \sim p(\theta)$ and nuisance parameters $\phi_i \sim p(\phi)$ and using a simulator to generate synthetic cryo-Em images $I_i \sim p(I|\theta=\theta_i, \phi=\phi_i)$. The simulated data $\mathcal{D} = \{(\theta_i, I_i)\}_{i=1}^N$ contains only pairs of images $I$ and conformations parameters $\theta$, and implicitly marginalize over $\phi$. cryoSBI trains a surrogate model on $\mathcal{D}$ to approximate the posterior $p(\theta| I)$. The surrogate model contains an embedding network $S_{\psi}$, which compresses the images into a lower dimensional vector of features $v_i = S_{\psi}(I_i)$. The second stage is a density estimator, which learns the posterior density $q_{\nu}(\theta|S_{\psi}(I))$ given the image features.

The trained posterior surrogate can accurately identify conformations displayed in synthetic and experimental images (Fig \ref{fig:4}D). The width of the posterior quantifies the inference precision, which depends crucially on the signal-to-noise ratio and projection direction \cite{dingeldein2024amortized}. The inference is amortized, paving the way to analyze massive cryo-EM data sets to search for rare structural intermediate and scarcely populated states. 

\section{Challenges and perspectives}

The primary challenge for SBI is model misspecification. Inference is only accurate if the assumed model is a faithful approximation of the experimental data-generating process \cite{cannon2022investigating}. A misspecification between the two models results in inaccurate inference. Efforts to detect and ameliorate model-misspecification include the development of calibration datasets that map experimental observations to simulated data with known parameters \cite{wehenkel2024addressing}, leveraging manifold-learning techniques to ensure that simulated and experimental data are close, and incorporating experimental data into training pipelines for embedding networks \cite{huang2024learning}. 

For complex models, simulations might be too expensive to generate enough data to train accurate surrogate models \cite{hermans2021trust}. Active learning strategies---where adaptive rounds of simulations gradually focus on the most probable regions of the likelihood or posterior---significantly reduce the number of required simulations \cite{greenberg2019automatic, deistler2022truncated, ahmed2021refinement, sharrock2022sequential, wiqvist2021sequential}.
Also, gradients and un-marginalized likelihoods evaluated in the simulator can be leveraged to accurately train surrogate models even if simulation data is sparse \cite{brehmer2020mining}. 

SBI is already a key technology in many scientific fields \cite{cranmer2020frontier}, enabling Bayesian inference with complex models, opening up new opportunities for data analysis, and influencing the types of data we acquire and the models we use to interpret them. The application of SBI for the analysis of single-molecule data is still in its infancy, but recent applications, particularly in smFS \cite{dingeldein2023simulation} and cryo-EM \cite{dingeldein2024amortized}, underscore its potential. SBI is a general framework that can easily be extended to analyze other single-molecule data, particularly using models with intractable likelihoods. 
SBI benefits from an active community \cite{SimulationbasedInferencea}, which develops and maintains powerful and easy-to-use open access code \cite{tejero-cantero2020sbi, sbidev}.

In the era of advanced computer simulations, SBI allows building on the power of detailed simulators as the current manifestation of scientific models for making principled statistical inference of complex experimental data.

\section*{Acknowledgments}

L.D. and R.C. acknowledge the support of Goethe University Frankfurt, the Frankfurt Institute of Advanced Studies, the LOEWE Center for Multiscale Modelling in Life Sciences of the state of Hesse, the CRC 1507: Membrane-associated Protein Assemblies, Machineries, and Supercomplexes (P09), and the International Max Planck Research School on Cellular Biophysics. P.C. acknowledges support by the The Flatiron Institute, a division of the Simons Foundation.
L.D. and R.C. thank the Flatiron Institute for hospitality.

%% If you have bibdatabase file and want bibtex to generate the
%% bibitems, please use
%%
\bibliographystyle{elsarticle-num} 
\bibliography{arxiv_20241021}

%% else use the following coding to input the bibitems directly in the
%% TeX file.

% \begin{thebibliography}{00}

% %% \bibitem{label}
% %% Text of bibliographic item

% \bibitem{}

% \end{thebibliography}
\end{document}